\documentclass[sigconf,authorversion,pdftex]{acmart}
\pdfoutput=1

\def\eg{{\it e.g.}}

\def\etal{et al.}

\makeatletter
\@ifclasswith{acmart}{sigconf}
{
    \def\bfigure{\begin{figure*}[t]}
    \def\efigure{\end{figure*}}
    \def\btable{\begin{table*}[t]}
    \def\etable{\end{table*}}
}{
    \def\bfigure{\begin{figure}[t]}
    \def\efigure{\end{figure}}
    \def\btable{\begin{table}[t]}
    \def\etable{\end{table}}
}
\makeatother

\AtBeginDocument{%
  \providecommand\BibTeX{{%
    \normalfont B\kern-0.5em{\scshape i\kern-0.25em b}\kern-0.8em\TeX}}}

\setcopyright{acmcopyright}

\copyrightyear{2021}
\acmYear{2021}
\setcopyright{acmlicensed}\acmConference[IUI '21]{26th International
Conference on Intelligent User Interfaces}{April 14--17, 2021}{College Station,
TX, USA}
\acmBooktitle{26th International Conference on Intelligent User Interfaces (IUI
'21), April 14--17, 2021, College Station, TX, USA}
\acmPrice{15.00}
\acmDOI{10.1145/3397481.3450664}
\acmISBN{978-1-4503-8017-1/21/04}

\acmSubmissionID{1011}

\begin{document}

\title[GO-Finder: A Registration-Free Wearable System for Assisting Users in Finding Lost Objects]{GO-Finder: A Registration-Free Wearable System for Assisting Users in Finding Lost Objects via Hand-Held Object Discovery}

\author{Takuma Yagi}
\email{tyagi@iis.u-tokyo.ac.jp}
\affiliation{%
  \institution{The University of Tokyo}
  \city{Tokyo}
  \country{Japan}
}

\author{Takumi Nishiyasu}
\email{nisiyasu@iis.u-tokyo.ac.jp}
\affiliation{%
  \institution{The University of Tokyo}
  \city{Tokyo}
  \country{Japan}
}
 
\author{Kunimasa Kawasaki}
\authornote{This research was conducted independently of Fujitsu (Laboratory), and the results are not representative of Fujitsu.}
\email{kunimasa.kawasaki@gmail.com}
\affiliation{%
  \institution{FUJITSU LABORATORIES LTD.}
  \city{Kanagawa}
  \country{Japan}
}

\author{Moe Matsuki}
\authornote{This research was conducted independently of SoftBank Corp., and the results are not representative of SoftBank.}
\email{matsuki.sousisu@gmail.com}
\affiliation{%
\institution{SoftBank Corp.}
  \city{Tokyo}
  \country{Japan}
}

\author{Yoichi Sato}
\email{ysato@iis.u-tokyo.ac.jp}
\affiliation{%
\institution{The University of Tokyo}
  \city{Tokyo}
  \country{Japan}
}

\begin{abstract}
People spend an enormous amount of time and effort looking for lost objects.
To help remind people of the location of lost objects, various computational systems that provide information on their locations have been developed.
However, prior systems for assisting people in finding objects require users to register the target objects in advance.
This requirement imposes a cumbersome burden on the users, and the system cannot help remind them of unexpectedly lost objects.
We propose GO-Finder (``Generic Object Finder''), a registration-free wearable camera based system for assisting people in finding an arbitrary number of objects based on two key features: automatic discovery of hand-held objects and image-based candidate selection.
Given a video taken from a wearable camera, Go-Finder automatically detects and groups hand-held objects to form a visual timeline of the objects.
Users can retrieve the last appearance of the object by browsing the timeline through a smartphone app.
We conducted a user study to investigate how users benefit from using GO-Finder and confirmed improved accuracy and reduced mental load regarding the object search task by providing clear visual cues on object locations.
\end{abstract}



\maketitle

\section{Introduction}\label{sec:introduction}

Looking for an object we do not remember leaving somewhere occurs frequently and is considered as a recurring problem regardless of age~\cite{peters2004finding}.
One survey reported that people waste 2.5 days a year looking for misplaced objects~\cite{pixiefoundsurvey}.
Thus, technological support to assist users in finding lost objects is demanded~\cite{eldridge1992memory,peters2004finding}.

Ubiquitous computing tackles this problem by collecting and providing cues on where objects are located.
Placing external sensors on the target object~\cite{shinnishi1999hide,kawamura2007ubiquitous}, and detecting objects with visual sensors~\cite{ueoka2003m,borriello2004reminding,xie2008design} are proposed as major solutions to keep track of object locations on behalf of users.
Such prior systems are designed to track a small number of important objects and ask a user to register target objects in advance to track those objects.
When looking for an object, the user searches a list of the registered objects (\eg, a list of object names) to select which object to look for.

However, objects we lose are not necessarily registered.
We often lose unique objects such as important documents or a new item bought the day before.
Since such objects are not usually registered, the system cannot help users find them.
To deal with such losses, we may think of to automatically registering all the objects appearing around the user.
However, this produces an enormous amount of candidates, which makes it impossible the user to find an object within a reasonable amount of time.
Moreover, assigning a unique name to each object will be unrealistic as the number of objects grows.
To support finding arbitrary objects, we need not only to track potential objects to be lost but also to eliminate the burden of registration.

\bfigure
\centerline{\includegraphics[width=0.9\linewidth]{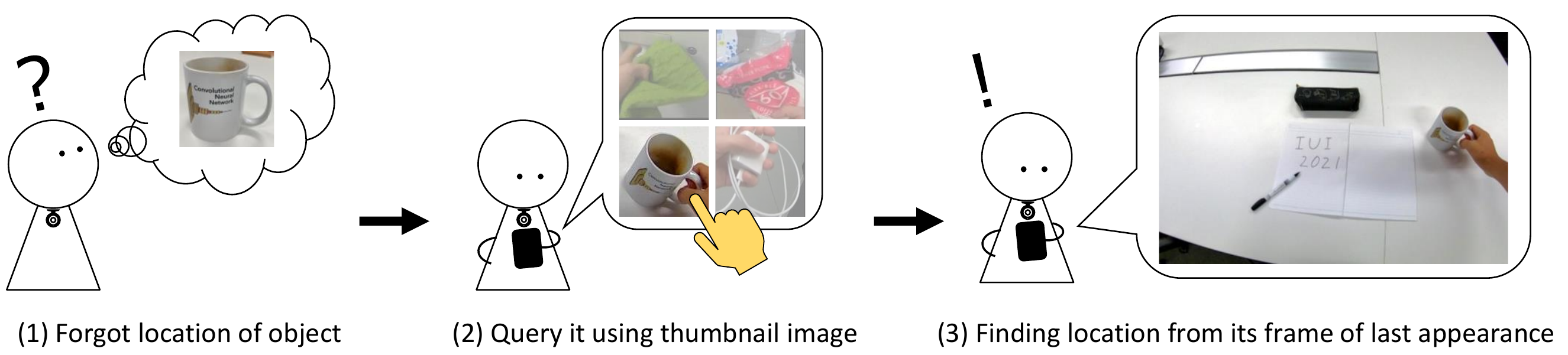}}
\caption{GO-Finder assists users in finding lost objects by showing last scene when the user handled it. User looks through list of object images to select object of interest.}
\label{fig:teaser}
\efigure

We introduce two key ideas to overcome these issues.
First, instead of tracking all the objects appearing around the user, we limit the search range to \emph{objects handled by hands}.
Since most portable objects we want to look for are handled with our hands, we can significantly reduce the number of candidate objects by limiting the scope to hand-held objects.
A reduced number of candidates enables users to look for the target object in a realistic amount of time.

Another key idea is to \emph{use the object image as a query} to select which objects to look for from the candidates.
Instead of assigning unique names to objects, we display to users a list of object images to select which object to look for.
Visual information of objects enables the user to identify the target object instantly without assigning a unique name to it.

Based on these ideas, we propose GO-Finder (``Generic Object Finder''), a registration-free wearable system for assisting users in finding arbitrary hand-held objects.
GO-Finder only requires a video captured from the wearable camera, does not require any registration, and can handle arbitrary hand-held objects automatically.
When finding objects by GO-Finder, users first skim through a list of object thumbnails, called \emph{the hand-held object timeline}, to ask the system which objects to search for.
Given the selected object, GO-Finder presents an image of the last scene when it appeared (Figure ~\ref{fig:teaser}).
This is achieved by a fully automatic process of \emph{hand-held object discovery}, which detects and clusters hand-held objects.

To validate the effectiveness of GO-Finder, we conducted a user study in a laboratory setting, mimicking a situation of finding an object.
We confirm that users can successfully find objects by using GO-Finder and reduce their mental load on performing the object-search task compared to the unaided condition.
Participant feedback suggests that it is feasible to find arbitrary hand-held objects using our proposed hand-held object timeline, which significantly broadens the coverage of objects to look for.

\section{Related Works}\label{sec:related}

\subsection{Computational Systems for Finding Lost Objects}
Various types of sensors, such as wireless tags~\cite{orr1999system,borriello2004reminding,liu2006ferret,wilson2007utilizing,tanbo2017active}, Bluetooth~\cite{kientz2006s,pei2010inquiry,schwarz2014cosero}, stationary cameras~\cite{butz2004searchlight,xie2008design}, and wearable cameras~\cite{ueoka2003m,funk2013antonius,funk2014representing}, have been studied for systems to assist users in finding lost objects.
Active and passive radio-frequency identification (RFID) tags are frequently deployed by attaching them to target objects.
While RFID tags are effective in indoor environments, they cannot locate an object when taken outside the search range.
To expend the search range, a combination of Bluetooth and GNSS are adopted in some commercial products (\eg, Tile~\cite{tile}).
Although these systems can provide the angle and distance from the tag, their guidance is less intuitive and attaching an external tag to each object will be a major bottleneck to track a large number of objects.

Alternatively, camera-based systems have the merit of not requiring external sensors attached to objects.
Butz \etal~\cite{butz2004searchlight} used augmented reality (AR) markers to search for objects in an office environment.
Xie \etal~\cite{xie2008design} proposed a dual-camera system for indoor object retrieval.
However, stationary cameras do not solve the problem of the search range and are weak against occlusions when objects are hidden by other entities.

Wearable camera-based systems mitigates these problems by capturing images from the user's viewpoint.
Since the camera moves along with the user, the system captures a close-up of the surrounding environments and it can be carried, significantly expanding the search range.
Similar to GO-Finder, Ueoka \etal~\cite{ueoka2003m} developed a wearable camera based object retrieval system based on object detection.
The system consists of head-mounted RGB and infrared cameras for capturing pre-registered objects.
It assists in object search by showing the last scene of the target object detected.
The same strategy is adopted in this work.
But unlike \cite{ueoka2003m}, our wearable-camera-based system, however, automatically groups all the hand-held objects appearing around the user, eliminating the registration operation.

Different from all the above works assuming a small number of items to be manually registered, we tackle the challenging problem of fully-automatic hand-held object tracking.
We provide the users with how to select the objects of interest efficiently from a list of automatically tracked objects.

\subsection{Camera-based systems for Mitigating Memory Problem}
Camera-based systems are used for mitigating memory problems other than losing objects since visual information offers a large amount of information better than textual information~\cite{brewer1988qualitative}.
Tran \etal~\cite{tran2005cook} proposed a system to monitor the progress of a cooking activity.
Hodges \etal~\cite{hodges2006sensecam} proposed a wearable camera based system called SenseCam, which takes wide-angle pictures periodically (\eg, one shot every 30\,s) to remind users of past events.
Li \etal~\cite{li2019fmt} proposed FMT, a wearable memory-assistance system to remember the state of objects (\eg, the last time the plant was watered).
While their hardware configuration is similar to ours in using neck-mounted wearable cameras, they aim to recall past interactions of a few numbers of daily-used objects, asking users to attach AR markers to each object.
In contrast, GO-Finder aims to expand the range of objects which could be searched for by removing the registration operation.

\subsection{Objects and Hands in First-Person Videos}

GO-Finder executes hand-held-object detection and grouping to discover objects appearing in first-person videos.
Discovering objects in first-person videos is a difficult problem since object categories appearing in daily life are massive, diverse, and individual-dependent.
To this end, various methods have been proposed to discover objects in first-person videos~\cite{lee2012discovering,bolanos2015ego,reyes2016my,bertasius2017unsupervised}.
Lee \etal~\cite{lee2012discovering} developed a model to discover important object regions using multiple first-person saliency cues.
Lu \etal~\cite{lu2015personal} proposed an object clustering-based method for personal-object discovery.
Their system involves object-scene distribution based on the assumption that personal objects appear in different scenes while non-personal objects typically remain in similar scenes.

Since objects appearing in first-person videos are typically handled by hands, hand information is used to improve object detection.
Lee \etal~\cite{lee2019hands,lee2020hand} proposed using hands as a guide to identify an object of interest from a photo taken by people with visual impairment.
Shan \etal~\cite{shan2020understanding} collected a large-scale dataset of hand-object interaction along with annotated bounding boxes of hands and objects in contact with each other.
Their proposed system can detect hands and objects in contact with each other from an image.
Our aim is not only detecting hand-held objects but also to discover hand-held-object instances from first-person videos, which reduces the number of candidates to be registered.

\section{System Design}\label{sec:interface}

\bfigure
\begin{tabular}{c}
\begin{minipage}{0.475\hsize}
\centerline{\includegraphics[width=0.95\linewidth]{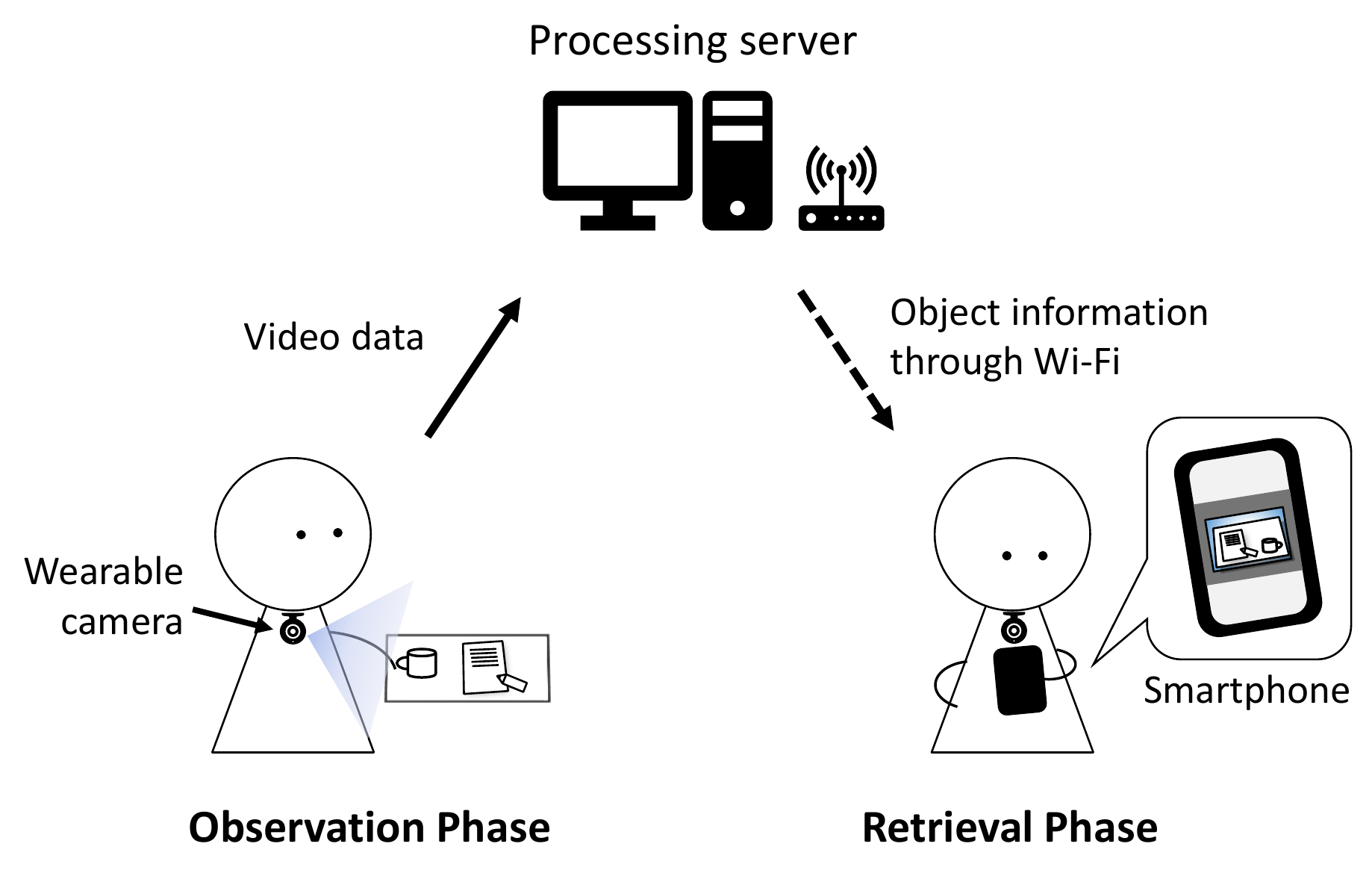}}
\captionsetup{width=0.9\textwidth}
\caption{Users wears wearable camera on their neck. During observation phase, their first-person images are sent to processing server to discover hand-held objects. At retrieval, processed results are sent from server, and user retrieves last frame of objects through smartphone app.}
\label{fig:setup}
\end{minipage}
\begin{minipage}{0.475\hsize}
\begin{center}
\centerline{\includegraphics[width=0.95\linewidth]{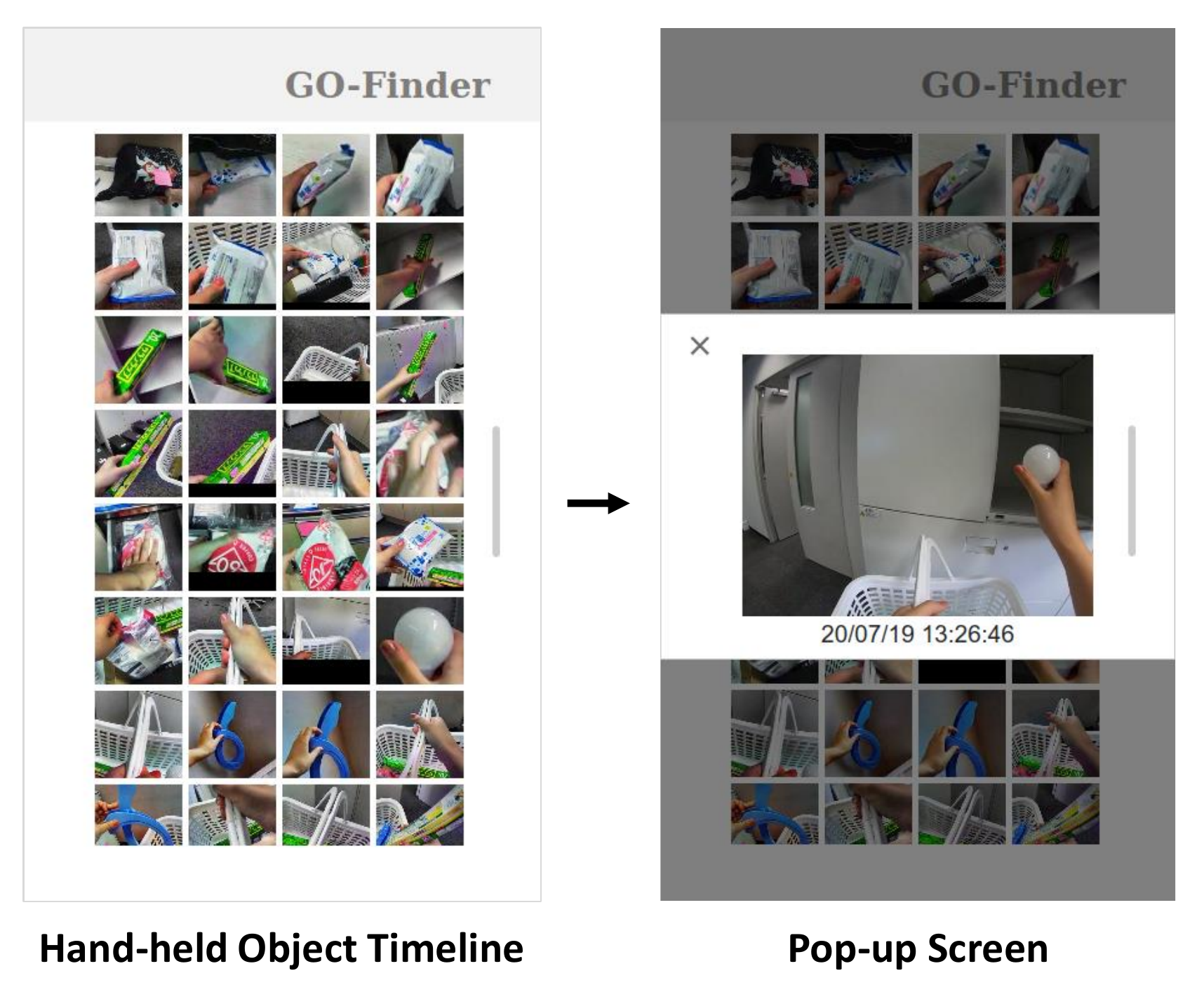}}
\captionsetup{width=0.9\textwidth}
\caption{Interface of smartphone app. (Left) Hand-held-object timeline. (Right) Pop-up screen.}
\label{fig:interface}
\end{center}
\end{minipage}
\end{tabular}
\efigure

\subsection{System Overview}
GO-Finder requires a wearable camera, processing server, and smartphone for browsing the location of objects the user is looking for (see Figure \ref{fig:setup}).
The procedure is divided into observation and retrieval phases.

In the observation phase, a user wears a camera on their neck.
The camera continuously stores the first-person images send to the processing server.
The server processes the received images to detect and track hand-held objects.
Finally, images are clustered by their appearance to discover groups of object instances.

In the retrieval phase, users use a smartphone-based interface (see Figure~\ref{fig:interface}) to receive the processed results thorough a wireless connection.
First, users select which object to look for through the hand-held object timeline (Figure~\ref{fig:interface} left).
Then, they find the target object by viewing the pop-up screen showing the last appearance of it (Figure~\ref{fig:interface} right).

\subsection{Hand-held Object Discovery}
GO-Finder attempts to detect hand-held objects and discover groups of object instances from the first-person video.
By discovering object instances, we can acquire the last appearance of the object, which is used to find the object.
Figure~\ref{fig:timeline} shows a rough sketch of how to acquire the last appearance of an object.
An object detector detects hand-held objects from first-person video frames.
From all the detected object images, we apply tracking and clustering (see Section~\ref{sec:algorithm} for details) to discover groups of cropped object images, clustered by instance.
Since we are interested in finding the last location of the object, we only use the last thumbnail image and last frame for our user interface.

\bfigure
\centerline{\includegraphics[width=0.9\linewidth]{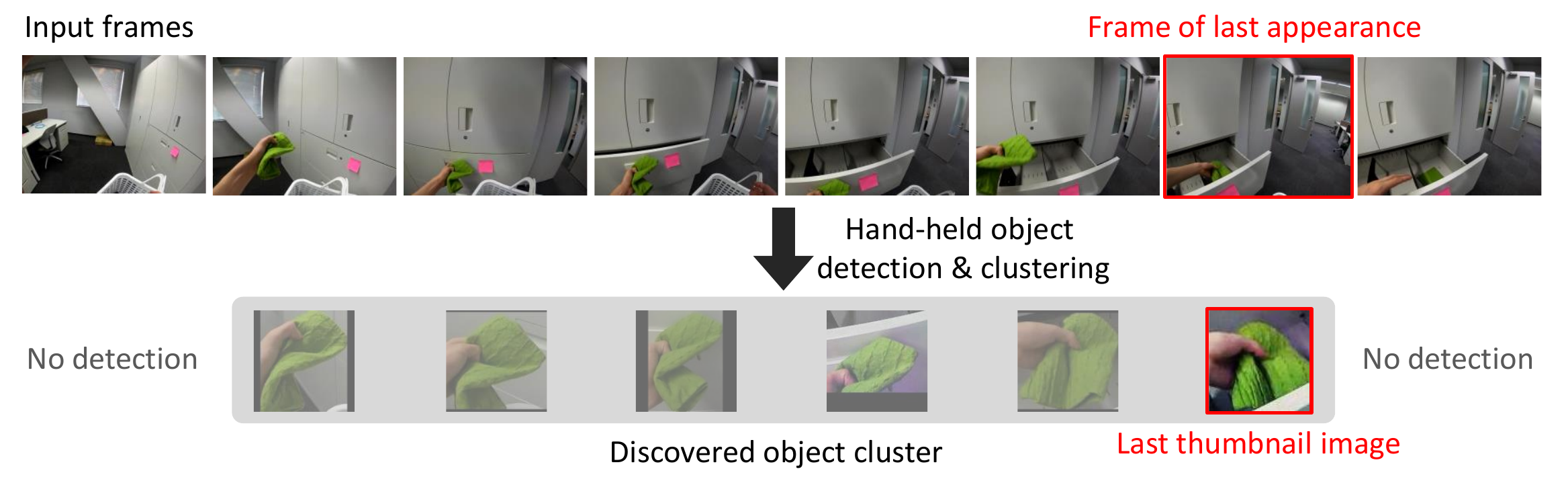}}
\caption{Given first-person video frames, system detects hand-held objects and groups them to discover cluster of cropped object images for each object. Since we are interested in providing last location where the object appeared, we use last thumbnail image and last frame in which the object appeared to help user find specific object.}
\label{fig:timeline}
\efigure

\subsection{Hand-held Object Timeline} 
GO-Finder automatically discovers hand-held object instances and registers them as candidates.
In this case, searching for objects by their names becomes unrealistic since it requires an association between the object name and its appearance.
We propose \emph{the hand-held object timeline}, which selects the target object by browsing the thumbnail images of the objects (see Figure~\ref{fig:interface} left). 
Thumbnails of the objects are sorted by the last time they appeared in descending order.
By skimming through the timeline, users select a thumbnail of the target object to retrieve its last appearance.
We adopt the image timeline as a metaphor for a photo album, which is widely accepted in existing smartphone-based interfaces.

Note that the obtained object timeline can be used as a trigger to remind the user of the object location.
The timeline acts as a concise history of what the user has handled in the past.
Even before arriving at the target object, the user can be reminded of past actions by looking back at the timeline.

\subsection{Pop-up Screen}
By clicking on a thumbnail of the object timeline, a pop-up screen will appear to show the appearance of the object and time (see Figure~\ref{fig:interface} right).
Since the pop-up screen shows the critical moment of leaving an object, the user can instantly be reminded of the location of the object by looking at the surrounding environments.

\section{Algorithm and Implementation}\label{sec:algorithm}
We introduce the details on the hand-held object discovery algorithm used in GO-Finder (see Figure~\ref{fig:algorithm}).

\bfigure
\centerline{\includegraphics[width=1.0\linewidth]{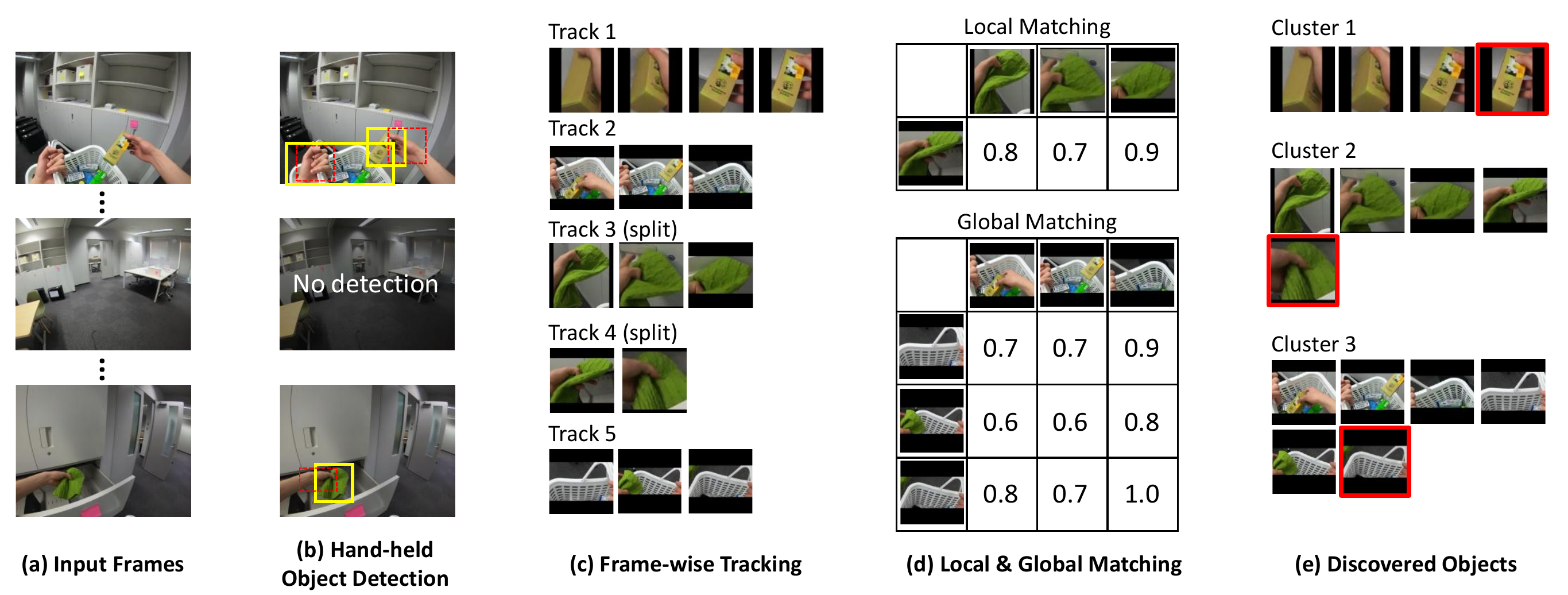}}
\caption{Overview of hand-held-object-discovery algorithm. (a) Input frames. (b) Example of hand-held object detection. Yellow and red boxes denote detected objects and hands, respectively. (c) Tracked detections. Typically, they are segmented due to tracking failure or re-appearance. (d) Local matching between latest detection and existing cluster (top). Global matching between two existing clusters (bottom). (e) Segments are clustered by instance. Last appeared scenes (images with red frame) will be displayed in user interface.}
\label{fig:algorithm}
\efigure

\subsection{Hand-held Object Detection}
We use the state-of-the-art algorithm on hand-held object detection~\cite{shan2020understanding} trained on a large-scale image dataset of hand-object interaction collected from first-person video datasets~\cite{sigurdsson2018actor,damen2018scaling,li2018eye}.
Given a video frame, it produces bounding boxes of the hand, contact state (self-contact, other people, portable object, and static object), and its manipulating objects (see Figure~\ref{fig:algorithm} (a)).
It detects arbitrary types of objects in contact with hands while rejecting other objects not handled by them.
Since we are interested only in portable objects, we extract detections that are predicted as a portable object in the contact state prediction.
Furthermore, detections that occupy more than half the side length of the frame are considered noise and are excluded from prediction.

\subsection{Object Instance Discovery}
Using the detected bounding boxes, we cluster them into a set of instances based on their appearance features.
Every detection should be assigned to a single cluster, and re-appearing objects should be merged into existing clusters.
To this end, we adopt a combination of local and global matching, which consists of three stages.

\subsubsection*{Stage 1: Frame-wise Tracking}
We first apply a visual tracker to the detected hand-held objects.
If the tracker successfully associates between consecutive detections, we assign the detection to the same cluster as the previous one (Figure~\ref{fig:algorithm} (c)).
Since first-person videos include large camera motion, we use an appearance-based tracker~\cite{bertinetto2016fully}, which performs similarity matching.
The cost assignment matrix is calculated by the intersection-of-union between all the tracker's predictions and actual detections.
Optimal assignment is achieved by using the Hungarian algorithm~\cite{kuhn1955hungarian}.

\subsubsection*{Stage 2: Local Feature Matching}
When the tracking fails, we apply local matching between the latest detection and existing clusters based on the object's appearance.
We use pre-trained convolutional neural network (CNN) features to find similar objects in the existing clusters.
For every detection, a 2048-dimensional feature vector is first extracted from the layer before the final layer of ImageNet-pretrained ResNet-50\,\cite{he2016deep}.
We then calculate the cosine similarity between the new detection and all the detections in the cluster (see Figure~\ref{fig:algorithm} (d), top).
Next, for each cluster, if the maximum and median of the similarity scores are above certain thresholds, the new detection is merged with that cluster.
We check the median score to avoid false associations.
If none of the clusters meets the condition, then a new cluster is created.

\subsubsection*{Stage 3: Global Cluster-wise Merging}
Since the previous stage matches against a single detection, it tends to form a new cluster if the viewpoint or boundary of the latest detection fluctuates even it should be merged.
To deal with such incorrectly segmented clusters, we try to merge clusters by global cluster-wise merging.
Given a pair of clusters, we calculate sample-wise cosine similarity between clusters, forming a similarity matrix (see Figure~\ref{fig:algorithm} (d), bottom).
Note that the scores calculated at stage 2 can be reused in this stage.
If the maximum and median of the similarity matrix exceed certain thresholds, the two clusters are merged.

However, this merging process is time-consuming, and should not be repeated every time.
To reduce the number of trials, we re-try merging only if the number of similarity matrix elements is more than two times that of the last trial.

\subsubsection*{Determining Similarity Thresholds}
Changing the hyperparameters (maximum and median similarity threshold) may affect user experience.
Stricter thresholds produce oversegmented and increased number of clusters while achieving higher recall on discovered target objects.
This makes it more difficult for the user to select the object of interest from the candidates.
In contrast, looser thresholds result in a smaller number of clusters with the risk of missing objects due to wrong associations.
A reduced number of clusters may make it easier for the user to select the target object, but it may be impossible to find it if it is incorrectly merged with other objects.
While we empirically selected these parameters during the study, we further introduce additional heuristics to explicitly suppress false associations.

\subsubsection*{Constrained Clustering using First-person Cues}
During similarity calculation, the hand-held object discovery algorithm sometimes shows a high similarity to a different object due to the appearance of the hand and similar textures, producing false associations.
Therefore, we introduce several heuristics to suppress such false associations.
If a detected bounding box or a pair of them satisfies the following conditions, the similarity of that pair is set to zero.

\begin{itemize}
\item {\bf Aspect ratio between two boxes:} if the ratio of the two bounding box aspect ratios is larger than 1.5
\item {\bf Ratio of skin color:} if the ratio of the skin-colored region (calculated using color histogram) is larger than 0.3
\item {\bf Area ratio of the object to the corresponding hand:} if the ratio of the two area ratios (area of the object bounding box to that of the hand bounding box) is larger than 1.5
\end{itemize}

\subsubsection*{Implementation Details}
We sampled video frames at 10\,fps, and further resized them into VGA resolution before processing.
While a smaller frame rate is enough to capture the timing of leaving an object, we find that a higher frame rate is better to track objects stably.
We set the maximum threshold to 0.8 and median threshold to 0.7.

\section{Evaluation Study}\label{sec:evaluation}
We conducted an in-lab experiment to determine (i) whether GO-Finder can correctly discover hand-held objects from the video and (ii) whether users can use the system to find target objects.
We hypothesized that by using GO-Finder, users can find objects correctly and quickly with less mental load.

\bfigure
\centerline{\includegraphics[width=1.0\linewidth]{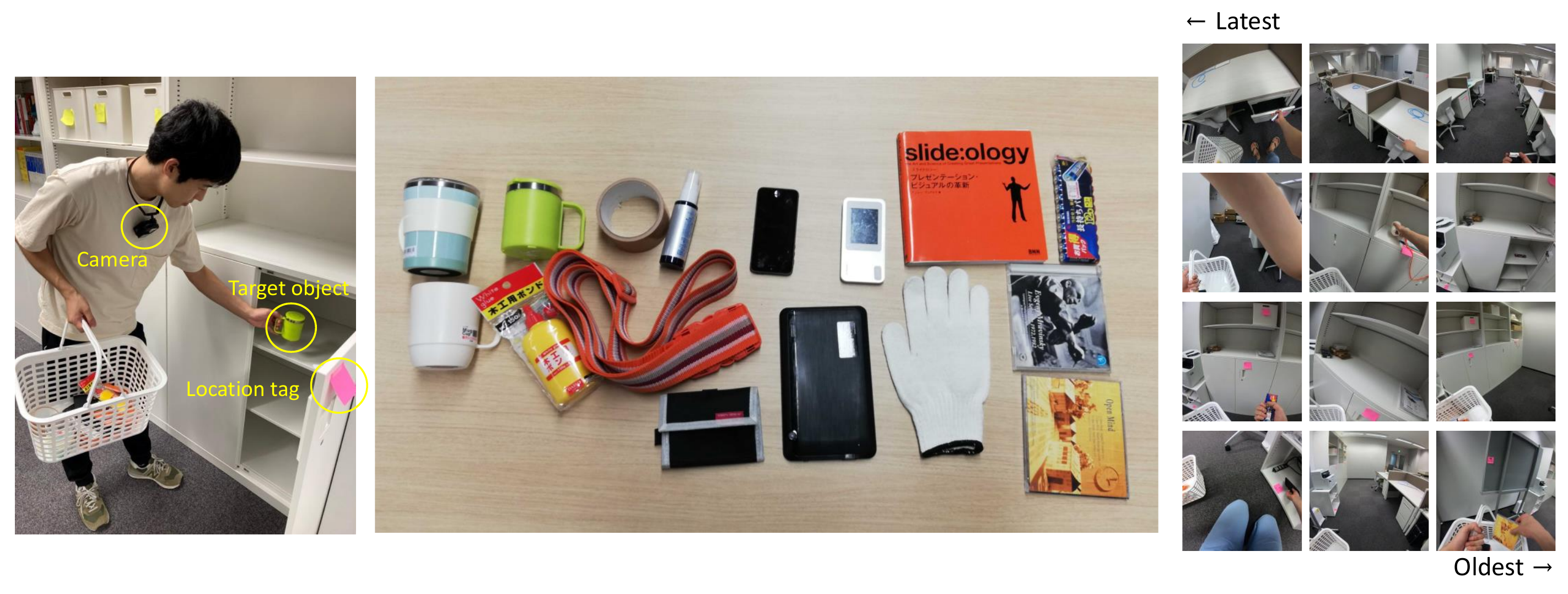}}
\caption{(Left) Object arrangement: participants hide objects at specified locations while wearing camera around their neck. (Middle) Objects used in study. (Right) Example of timeline of frame-based system.}
\label{fig:arrangement}
\efigure

\subsection{Procedure}\label{section:evaluation_1}
We recruited 12 volunteers\footnote{One participant (P05) was excluded from the analysis due to a misunderstanding of instruction.} (10 males and 2 females) with ages ranging from 18 to 28.
They were all familiar with using smartphones.
The experiment was conducted in a room in our lab.
The task was a \emph{hide-and-seek} task performed by the participants.
First, participants filled out a pre-study questionnaire on their past experience of looking for lost objects.
After an introduction to the task, each participant was asked to hide a set of objects inside a room (arrangement phase), conduct a surrogate task to forget the locations of the objects (forgetting phase), and later asked to correctly retrieve a subset of them (retrieval phase).
The trial was repeated three times, changing the experimental conditions.
Conditions were randomly shuffled to eliminate order effects.
After all the trials, participants filled out a post-study questionnaire on the usability of the interface. 
Finally, we conducted a semi-structured interview to find further insights.

\subsubsection*{Arrangement Phase}
First, the participant went to the room and asked to hide a set of objects prepared by the experimenter.
The locations to hide the objects were specified with pink tags and the participants were informed about them in advance (see Figure~\ref{fig:arrangement} left).
The participants carried a basket along with the objects.
During the experiment, participants wore a GoPro HERO 7 camera ($150^\circ$ diagonal field-of-view) to record first-person videos.

\subsubsection*{Forgetting Phase}
The participants moved to another room for a 15-min interval to forget the arrangements.
During the interval, the participant was asked to solve as many of a series of simple calculation problems as possible.

\subsubsection*{Retrieval Phase}
The participants came back to the room and were asked to bring back a subset of the hidden objects.
The list of objects to bring back was shown in a photo.
In addition to the neck-mounted camera, the participants wore smartphones around their necks to use the system.
Under each condition, participants were given instructions on how to use the system and become familiarized with the interface by browsing the result of a sample video carrying a few objects.
They were not forced to use the system; they used the system only when they needed to use it.

\subsection{Experimental Conditions}\label{section:evaluation_2}

We compared three conditions:
\begin{itemize}
    \item {\bf No aid}: The participant search for objects themselves without any assistance.
    \item {\bf Frame-based aid}: The participant is shown a timeline of images extracted every 5\,sec.
    \item {\bf Object-based aid (GO-Finder)}: Our proposed system with hand-held object timeline and pop-up screen.
\end{itemize}
The frame-based aid condition resembled automatic image capture devices such as SenseCam~\cite{hodges2006sensecam}.
We hypothesized that past images would help the participants remember their arrangement of objects.
Regarding the duration of the task, we showed images taken every 5\,sec (see Figure~\ref{fig:arrangement} right), which is denser than typical devices (\eg, 30\,sec).

We used a laptop PC to run the object discovery algorithm.
The connection between the laptop PC and smartphone was established via Wi-Fi as shown in Figure~\ref{fig:setup}.
At every trial, participants hid 16 objects in a choice of 20 locations and asked to retrieve 6 objects from them.
We used different object sets for each trial, resulting in 48 objects in total (see Figure~\ref{fig:arrangement} right).
The objects differed in color and shape, and sometimes included multiple instances of the same category.

\subsection{System Evaluation Measures}\label{section:evaluation_4}
\subsubsection*{Localization Rate}

To measure how well the hand-held object discovery algorithm can discover target objects, we counted the number of objects in which their locations are identifiable by a third person, who did not have any memory of arrangement; only using our system.
We define the localization rate as a ratio of the number of identifiable objects to the total number of target objects.
One of the authors manually calculated this metric by using the app.
We counted as a success only if a close-up of an object is visible in the thumbnail of the timeline and the object location could be correctly determined from the pop-up screen without difficulty.
This metric acts as an expected recall of the system.

\subsubsection*{Number of Clusters}
We also measured the number of clusters formed with the hand-held object discovery algorithm and analyzed the contents of the timeline. We ran the algorithm for all 36 trials (12 participants $\times$ 3 conditions).

\subsection{Objective Evaluation Measures}
\subsubsection*{Correctness of Retrieval}
We calculated the mean precision of each trial.
We counted as correct when the user found an object listed on the target list and incorrect when the user opened a location with the incorrect or no objects.
We compared three combinations of two of the conditions by using the paired t-test on the difference of mean scores.

\subsubsection*{Task Completion Time}
We expected shorter task completion time by using the system. 
We compared three combinations of two conditions by using the Wilcoxon signed-rank test on the difference in mean task completion times.

\subsubsection*{System Usage Time}
Since GO-Finder can search for objects directly through the hand-held object timeline, we expected to have a shorter usage time using GO-Finder to the frame-based aid condition.
We measured the number of times participants used the system\footnote{We counted as one time when the user attempted to search a location after using the system.} and usage time per trial from the recorded videos.

\subsection{Subjective Evaluation Measures}
\subsubsection*{Questionnaire}
After all the trials, participants answered questions on each condition.
First, participants were given the question, \emph{``How do you rate the difficulty of completing the task?"} on a seven-point scale (easy\,=\,1, difficult\,=\,7). We used the Wilcoxon signed-rank test in the difference of means.
Regarding the features of the interface, we asked whether they agreed to the following questions on a five-point scale: Q1) The timeline is easy to view. Q2) The timeline is intuitive to use. Q3) The timeline helped me look for objects. Q4) The pop-up screen is easy to view. Q5) The pop-up screen helped me look for objects. Q6) The timeline (under each condition) gave me a clue on the location of the target object. Q7) I could be reminded of the locations of the objects by using the system (under each condition).

\subsubsection*{Observation and Interview}
We observed how the participants searched for objects.
During the interviews, we asked what they thought during the retrieval task.
To collect insights on using this system in daily life, we also asked \emph{``What do you recommend to improve the interface?''}, and \emph{``How do you feel about wearing a camera in private/public places?''}.

\section{Results}\label{sec:results}
\begin{table}[t]
\caption{Localization rate of each object set (\%).}
\begin{center}
\scalebox{.95}{
\begin{tabular}{lrrr}
\hline
 & {\bf Mean $\pm$ SD} & {\bf Min} & {\bf Max} \\ \hline
Set 1 & 84.9 $\pm$ 7.8 & 68.8 & 93.8 \\
Set 2 & 83.3 $\pm$ 9.7 & 68.8 & 100 \\
Set 3 & 88.5 $\pm$ 10.9 & 62.5 & 100 \\ \hline
All sets & 85.6 $\pm$ 9.6 & 62.5 & 100 \\ \hline
\end{tabular}
}
\label{tab:discovery}
\end{center}
\end{table}

\bfigure
\centerline{\includegraphics[width=1.0\linewidth]{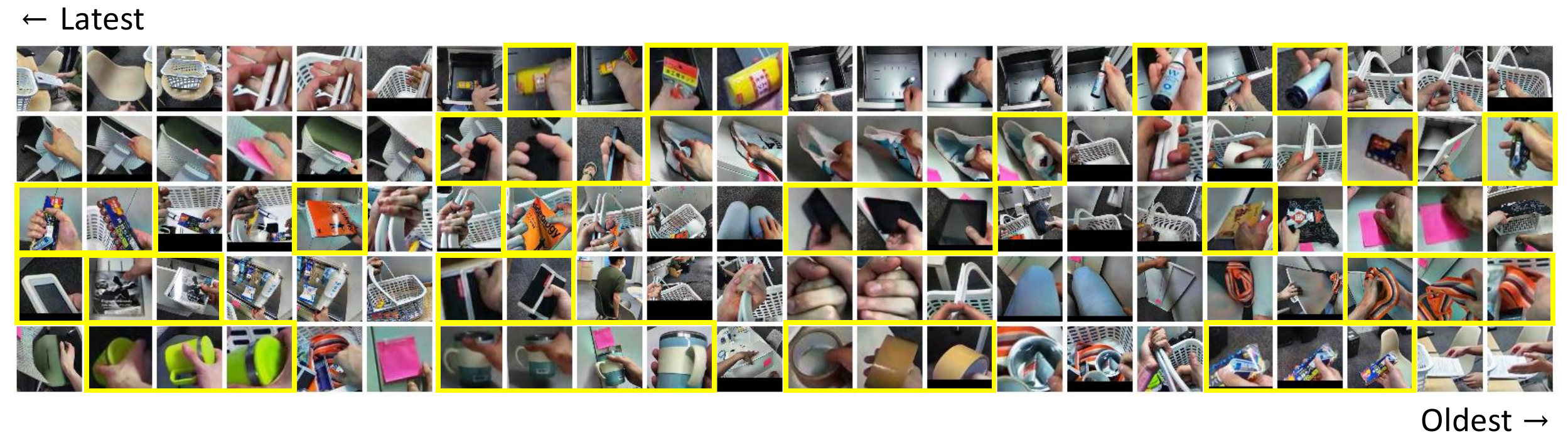}}
\caption{Example results of hand-held object timeline (localization rate=0.9375, \#clusters=110). Yellow boxes denote clusters that contain target objects. Some clusters were over-segmented into few clusters per object.}
\label{fig:timeline_08n}
\efigure

\subsection{System Evaluation}
\label{ssec:system_evaluation}

\subsubsection*{Localization Rate}
Table~\ref{tab:discovery} shows the localization rate of the hand-held object discovery algorithm.
The average score was 84.9, 83.3, and 88.5\% for each object set, and the overall average was 85.6\%.
These results indicate that GO-Finder can correctly display 13.6/16 objects per session on average.
The minimum rate across the participants were 62.5\%.

We found differences in performance among objects.
While several objects were discovered in all 12 trials (green cup, wood glue, electric bulb, futon pincher, green cloth, and teddy bear), some objects were difficult to discover (waiter's corkscrew: 16.7\%, medicine bottle: 58.3\%, black wallet, and spray bottle: 66.7\%).
We found that small and black objects were difficult to correctly discover due to occlusion and texture-less regions.

\subsubsection*{Number of Clusters}
The number of clusters (objects) that appeared in the hand-held object timeline was 108.6 ($SD=24.0$) on average.
Although it is not trivial to count the number of valid objects which should be discovered, the estimated number of valid objects (including furniture, drawers, and baskets) was expected to be about 20 to 30, including the 16 target objects.
Thus, we can conclude that the algorithm over-segments an object into 4 to 5 clusters on average.

\subsubsection*{Qualitative Analysis}
Figure~\ref{fig:timeline_08n} shows an example of the obtained hand-held object timeline. 
We annotated the thumbnail images that contain close-ups of the target objects in green boxes.
Forty out of 110 clusters contained 15 of the target objects.
In addition of the target objects, GO-Finder discovered various valid objects and false positives.
Examples of valid objects were chairs, baskets, and drawers while untouched furniture, participant's body, and other people were discovered as false positives.
While most objects were easily identifiable from the thumbnail images, some thumbnails were difficult to identify due to occlusion, shadow, and irregular views (\eg, the green cup in Figure~\ref{fig:timeline_08n} left-bottom).

\subsection{Object Retrieval Performance}

\btable
\begin{tabular}{cc}
\begin{minipage}{0.325\hsize}
\caption{Retrieval performance (precision).}
\begin{center}
\scalebox{.9}{
\begin{tabular}{lr}
\hline
 & {\bf Mean $\pm$ 95\% CI} \\ \hline
No aid & 0.728 $\pm$ 0.174 \\
Frame-based aid & 0.736 $\pm$ 0.124 \\
Object-based aid & 0.922 $\pm$ 0.084 \\ \hline
\end{tabular}
}
\label{tab:retrieval}
\end{center}
\end{minipage}
\begin{minipage}{0.625\hsize}
\caption{Results of paired t-tests on difference in mean precisions.}
\begin{center}
\scalebox{.85}{
\begin{tabular}{lrrrrr}
\hline
& & & \multicolumn{2}{c}{{\bf 95\% CI}} & {\bf Effect size}\\
& $t$ & $p$ & LB & UB & $d$ \\ \hline
No aid/frame-based aid & -0.104 & 0.918 & -0.193 & 0.175 & 0.04 \\
No aid/object based-aid & -2.012 & 0.069 & -0.407 & 0.018 & {\bf 0.95} \\
Frame-based aid/object-based aid & -2.339 & {\bf 0.039} & -0.360 & -0.011 & {\bf 1.17} \\ \hline
\end{tabular}
}
\label{tab:retrieval_test}
\end{center}
\end{minipage}
\end{tabular}
\etable

Tables~\ref{tab:retrieval} and \ref{tab:retrieval_test} show the results of the object retrieval task under each condition.
We report the average precision and its 95\% confidence interval (CI) under each condition.
As expected, GO-Finder showed better precision with less variance than the other two conditions.
The paired t-test revealed significance only between the frame-based aid and object-based aid conditions ($p=0.918$, $p=0.069$, and $p=0.039$, respectively).
However, both no aid/object-based aid and frame-based aid/object-based aid conditions showed large effect sizes ($d=0.95$ and $d=1.17$, respectively), indicating a positive effect by using the proposed system.
In contrast, we did not observe a marked difference between no aid and frame-based aid conditions ($d=0.04$).

\btable
\begin{tabular}{cc}
\begin{minipage}{0.325\hsize}
\caption{Task completion time (sec).}
\begin{center}
\scalebox{.9}{
\begin{tabular}{lr}
\hline
 & {\bf Mean \& 95\% CI} \\ \hline
No aid & 216 $\pm$ 142 \\
Frame-based aid & 238 $\pm$ 75 \\
Object-based aid & 178 $\pm$ 52 \\ \hline
\end{tabular}
}
\label{tab:completion}
\end{center}
\end{minipage}
\begin{minipage}{0.625\hsize}
\caption{Results of paired t-tests on difference in mean task completion times.}
\begin{center}
\scalebox{.9}{
\begin{tabular}{lrrrrr}
\hline
& & & \multicolumn{2}{c}{{\bf 95\% CI}} & {\bf Effect size}\\
& $t$ & $p$ & LB & UB & $d$ \\ \hline
No aid/frame-based aid & -0.316 & 0.379 & -173 & 130 & 0.12 \\
No aid/object-based aid & 0.513 & 0.309 & -128 & 206 & 0.23 \\
Frame-based aid/object-based aid & 1.530 & 0.077 & -27 & 148 & {\bf 0.60} \\ \hline
\end{tabular}
}
\label{tab:completion_test}
\end{center}
\end{minipage}
\end{tabular}
\etable

Table~\ref{tab:completion} and \ref{tab:completion_test} shows the result of the task completion time in each condition.
We did not observe improvement in task completion time by using GO-Finder.
The paired t-test did not show any significant difference ($p=0.379$, $p=0.309$, and $p=0.077$).
The average time and 95\% confidence interval of the arrangement phase was 223 $\pm$ 16\,sec.

\subsubsection*{Usage Time}
During the 12 sessions under the frame-based aid and object-based aid conditions, participants used the interface 32 and 35 times, respectively.
The mean (median) usage times were 28.1\,sec (23.0\,sec) and 16.1\,sec (12.5\,sec), respectively.
The paired t-test revealed a significant difference with medium effect size in the mean times ($p=0.005, d=0.71$).
This suggests that participants were able to browse the timeline more efficiently under the object-based aid condition than under the frame-based aid condition.

\subsection{Questionnaire}

\subsubsection*{Ease of Task}
Figure~\ref{fig:ease} and Table~\ref{tab:ease_test} show the results on ease of the task.
Surprisingly, the participants evaluated the frame-based aid condition the most difficult.
They evaluated the object-based aid condition the easiest among the three conditions.
Based on the Wilcoxon signed-rank test, we found a significant difference in the mean scores between the frame-based aid and object-based aid conditions ($p=0.063$, $p=0.133$, and $p=0.043$).
However, we observed medium effect size in all the combinations ($r=0.38$, $r=0.30$, and $r=0.41$).
This suggests that the participant's subjective mental load have decreased by using GO-Finder.

\subsubsection*{Functionality of Interface}
Figure~\ref{fig:questionnaire} shows the results of questions Q1--Q7.
In Q1--Q5, participants reported positive impressions with the proposed system.
The Wilcoxon signed-rank test revealed a significant difference between the frame-based aid and object-based aid conditions in Q6 ($p=0.007$) but not in Q7 ($p=0.065$).
However, Q6 and Q7 showed large and medium effect sizes ($r=0.55$, and $r=0.38$), respectively, suggesting that GO-Finder was more useful in finding object locations compared to under the frame-based aid condition.

\bfigure
\begin{tabular}{c}
\begin{minipage}{0.425\hsize}
\centerline{\includegraphics[width=1.0\linewidth]{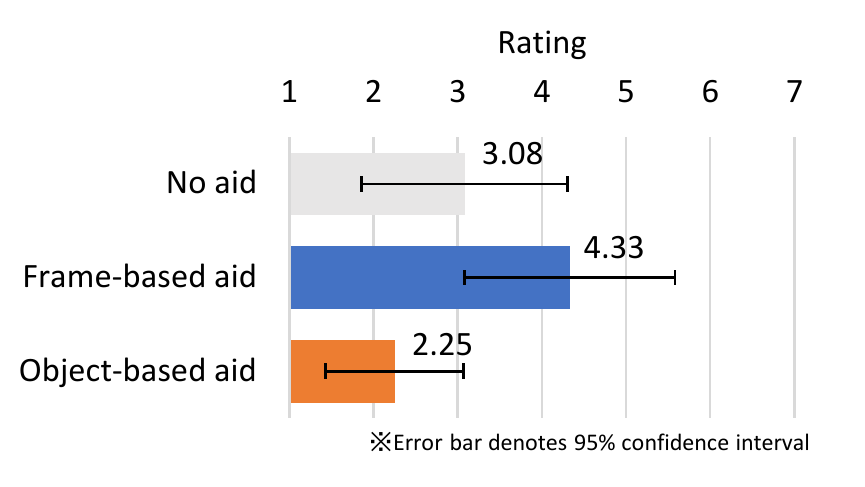}}
\caption{Ease of task (easy=1, difficult=7).}
\label{fig:ease}
\end{minipage}
\begin{minipage}{0.525\hsize}
\centerline{\includegraphics[width=1.0\linewidth]{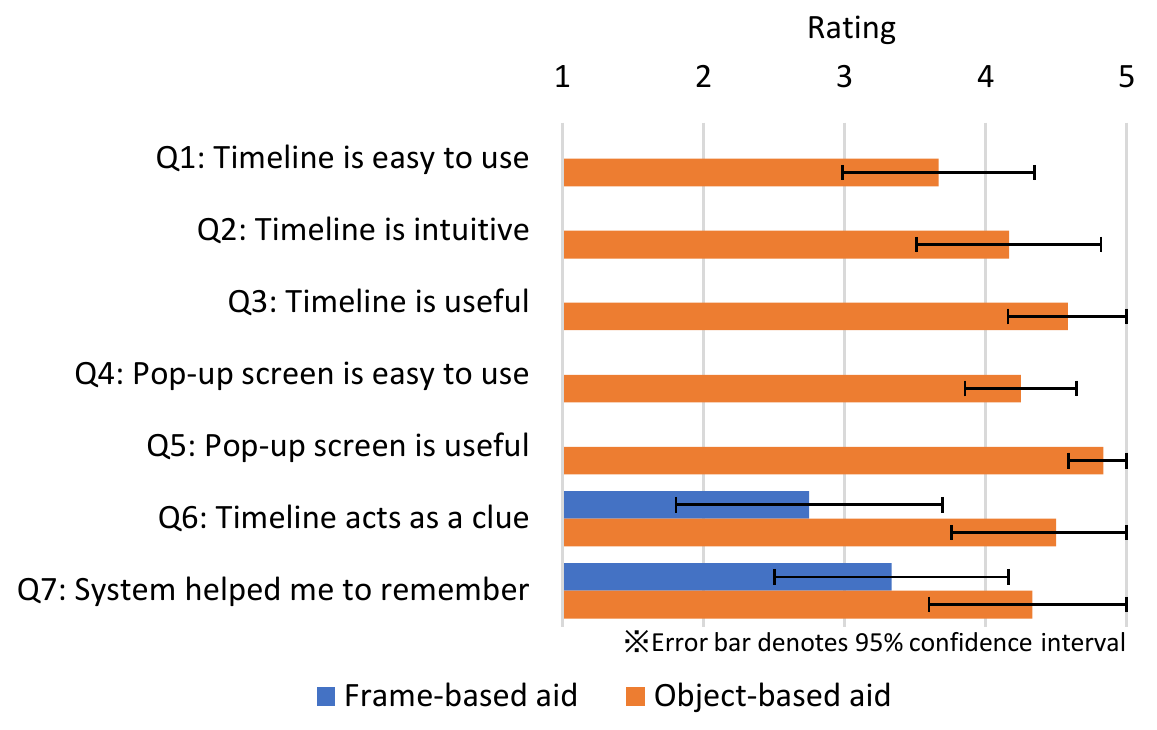}}
\caption{Results of questionnaire.}
\label{fig:questionnaire}
\end{minipage}
\end{tabular}
\efigure

\btable
\begin{center}
\caption{Result of Wilcoxon signed-rank tests on ease of task.}
\scalebox{.9}{
\begin{tabular}{lrrrrr}
\hline
& & & \multicolumn{2}{c}{{\bf 95\% CI}} & {\bf Effect size}\\
& $Z$ & $p$ & LB & UB & $r$ \\ \hline
No aid/frame-based aid & -1.857 & 0.063 & -2.581 & 0.081 & {\bf 0.38} \\
No aid/object-based aid & -1.501 & 0.133 & -0.596 & 2.263 & {\bf 0.30} \\
Frame-based aid/object-based aid & -2.027 & {\bf 0.043} & 0.540 & 3.627 & {\bf 0.41} \\ \hline
\end{tabular}
}
\label{tab:ease_test}
\end{center}
\etable

\subsection{Observation and Feedback}
\subsubsection*{Video Observation}
In general, participants first looked for objects that they remembered and used the system when they were not confident with the location.
When using the system, they looked for thumbnails showing the target object, inferred the location from the pop-up screen, and successfully retrieved the object.
Two users persist using GO-Finder even when the system failed to discover the target objects (P06, 10).

\subsubsection*{Usefulness of Object-based Aid Condition} %

Eleven out of the 12 participants said that GO-Finder was convenient to use.
Only with a brief instruction, they were able to retrieve the forgotten locations with GO-Finder.
They preferred the intuitiveness of the hand-held object timeline: ``{\it The function I wanted most was there. Because the objects were highlighted and zoomed in, I could see the target objects and retrieve their last appearance by tapping the thumbnail}'' (P08).

They felt more secure and confident at retrieval: ``{\it Since I don't have to rely on my intuition, I looked at the smartphone once I felt lost. By using the system, I often felt confident about the location}'' (P10).

A few participants trusted the system's output rather than their memory: ``{\it I arrived at the wrong location since I relied on the system. I didn't remember my memory but inferred the location from the pop-up screen and got wrong}'' (P13).

\subsubsection*{Comparison to Frame-based Aid} %
In contrast, nine participants gave negative feedback regarding the frame-based aid condition.
They mainly complained that the timeline often did not capture the exact moment of leaving objects.
Difficulty in finding critical scenes from large field-of-view images was also reported: ``{\it Since the images often don't capture the scene when holding objects, I found myself zooming into the image but found nothing several times}'' (P10).
The user has to additionally remember how they left the objects during the arrangement, sometimes being confused by their behavior: ``{\it I was deluded by myself attempting to leave the object once but actually done it afterward}"" (P03).

One participant preferred the frame-based timeline because thumbnails were evenly sampled in chronological order: ``{\it I preferred that (the frame-based timeline) because the entire timeline was available and I could infer how I searched by looking at an image and the image next to it}'' (P06).

\subsubsection*{On Interface of System}
Participants preferred thumbnail images given from their point of view: ``{\it The objects were shown by image, and were taken when I lost the object. The system was convenient since critical moments were captured in the timeline}'' (P09).
While participants gave positive feedback for every component of the interface (Q1--Q7), they gave lower scores on the ease of using the hand-held object timeline (Q1).
First, over-segmentation of the objects confused some participants: ``{\it Regarding four thumbnails showing a tennis ball, I had no idea which one to press[...]}'' (P01).
The quality of thumbnail images (brightness, occlusion, contrast, and viewpoint) also made it difficult for the participants to find the object of interest: ``{\it [...]the thumbnail image of the last scene was difficult to identify. For example, I had to zoom in (to the thumbnail) when I looked for the pouch}'' (P02).

\subsubsection*{Privacy Concerns}
While we expected to have negative feedback on capturing images, six participants reported that they were not concerned with recording videos while three participants raised specific concerns: ``{\it I don't feel any discomfort since I know what the system does; maybe because I know the system only collects information on objects. It might be different if the system captures people's faces}'' (P12).
Some participants changed their behavior since they were aware of being recorded even though we did not give them any warning: ``{\it I thought it was better not to hide the camera}'' (P02).

\subsubsection*{Suggestions on Improvement}
Two participants suggested playing a video snippet instead of a static image on the pop-up screen: ``{\it I think it'll be easier to remember if I can view the before and after of the last scene.}'' (P09).
Regarding real-world use, participants suggested querying by background (P01, 04) and time (P12).
They stated that the object itself is not a key to remember a scene and requested a functionality to filter the candidates by their own.

\section{Discussion}\label{sec:discussion}

\bfigure
\centerline{\includegraphics[width=0.65\linewidth]{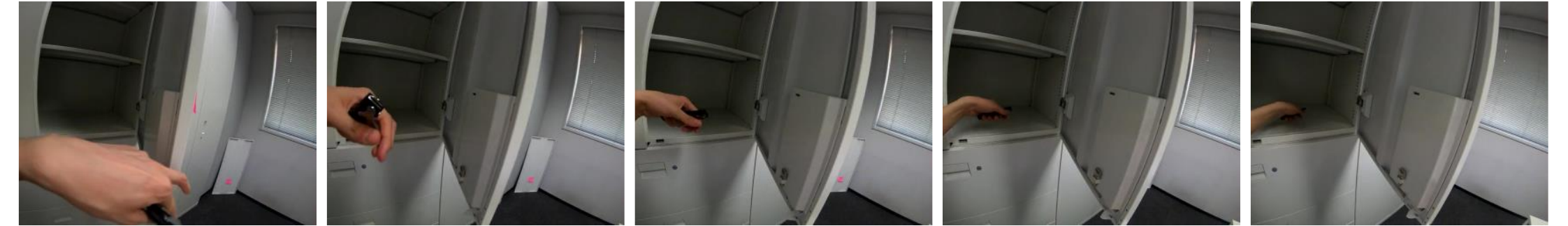}}
\caption{System failure under severe occlusion.}
\label{fig:failure}
\efigure

\subsection{Usefulness of GO-Finder}
In the user study, we confirmed that GO-Finder enabled the participants to retrieve hidden objects with less mental load.
Quantitative and qualitative feedback suggests that the users gained confidence by immediately accessing the last moment of when the objects were seen.
The frame-based timeline showing uniformly sampled video frames was not effective in this task.
Since the object finding task should be solved as quickly as possible, users requested direct access to the location rather than having to keep relying on their memory.

System evaluation showed that the hand-held object discovery algorithm of GO-Finder successfully extracted hand-held object instances while efficiently excluding other unrelated objects.
GO-Finder worked without explicit registration, making it easy for users to start using it.

Regarding the idea of using object images as a query, participants adapted quickly to browsing objects using the hand-held object timeline (Q3).
The timeline shown as a list of images was evaluated as intuitive and participants were able to access the object of interest immediately.
Since the thumbnail images are captured from almost the same view as the participant's one, this timeline also worked as a clue to remembering (Q6).

While most participants did not report any problem with querying by images of objects, some suggested using background and time information as additional cues for retrieval.
This feedback suggests that participants wanted to be reminded of the scene by using their incomplete memory when browsing target objects through the timeline.
In this study, we did not provide any features to narrow down the candidates by contextual information such as scene, location, or time.
Adding such features would provide options to users reaching the target object in their most familiar way.

\subsection{Privacy Issues}
The use of wearable cameras raises privacy concerns in real-world use~\cite{denning2014situ,hoyle2014privacy}.
Although GO-Finder's contents are not shared among other users, the privacy and comfort of bystanders must be secured.
One way out of the difficulty is to filter out sensitive contents while storing only the information relevant to hand-held objects.
Since GO-Finder only requires the last scene of an object, other frames are no longer needed as we store the feature vector of object detections.
Last scenes will be updated as objects re-appears so images would not be kept stored permanently.
Additionally, we can remove identity information by running an off-the-shelf face detector since we do not need bystander's information.
Supported by the positive comments, we believe that GO-Finder can be used with minimal interference.

\subsection{Limitations}

\subsubsection*{Object Re-identification}
The proposed system failed to discover objects under severe occlusion by hands (\eg, waiter's corkscrew, see Figure~\ref{fig:failure}).
In this example, the participant gripped the corkscrew so that it was severely occluded by the hand.
This confused both the detector and clustering algorithm in determining the correct bounding box and appearance feature to identify the object, resulting in over-segmentation or false cluster merge.
As reported in \ref{ssec:system_evaluation}, small objects tend to be occluded and would be problematic considering real-world use.
One potential solution is to use the object's appearance before or after but not during manipulation with occlusion by hands.

\subsubsection*{Long-term Evaluation}
We evaluated GO-Finder using short video sequences of around 4\,min.
Although we revealed that GO-Finder can handle around 20 objects without registration, in a real situation, the video length will be hours or days, and the number of objects will increase as we record more.
To avoid having to look over a massive amount of candidate objects, additional features to omit unimportant detections are necessary.
For instance, we can use object-scene distribution~\cite{lu2015personal} to eliminate non-personal, static objects such as doors and furniture.

\subsubsection*{Multi-user Scenarios}
We assume each object is manipulated by a single user. However, in multi-user scenarios, we cannot track objects if they are moved by other people.
One plausible solution is to share object information among users, which would be a trade-off between privacy protection.

\section{Conclusions}\label{sec:conclusions}
We presented GO-Finder, a registration-free wearable camera-based system for assisting users in finding lost objects.
It supports finding of arbitrary number of objects based on two key ideas: hand-held object discovery and image-based candidate selection.
The user study revealed that by using GO-Finder users can find the location of lost objects correctly with a reduced mental load.
Even the objects were registered automatically without user intervention, the users were able to identify the target object using the image-based hand-held object timeline.
Going beyond tracking only a few selected objects, GO-Finder could be used as a practical tool to help find various unexpectedly lost objects in daily life.

Our future work include developing a computationally efficient object discovery algorithm and candidate filtering based on contexts as well as conducting a long-term user evaluation on naturalistic situations of losing objects.

\section*{Acknowledgements}
This work was supported by JST AIP Acceleration Research Grant Number JPMJCR20U1 and Masason Foundation.

\newpage
\appendix
\section{Dataset Details}
\bfigure
\centerline{\includegraphics[width=1.0\linewidth]{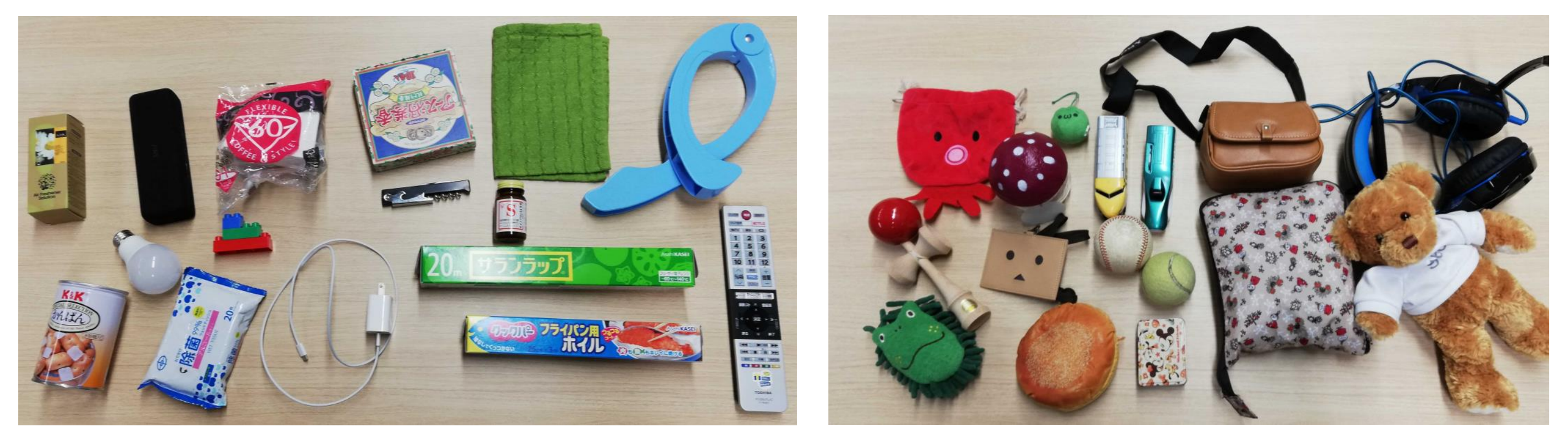}}
\caption{Other object sets used in user study.}
\label{fig:objects}
\efigure

Figure~\ref{fig:objects} shows the other two object sets (set 2 and set 3) used in the study.
To avoid confusion across object sets, we used different types of daily objects between trials.

\section{Additional Results}
\begin{table}[t]
\caption{Result of the SUS test.}
\begin{center}
\begin{tabular}{lrr}
\hline
{\bf ID} & {\bf Gender} & {\bf SUS score (rank)} \\ \hline
P01 & Female & 60 (D) \\
P02 & Male & 65 (C) \\
P03 & Female & 80 (A) \\
P04 & Male & 47.5 (F) \\
P06 & Male & 85 (C) \\
P07 & Male & 77.5 (B+) \\
P08 & Male & 77.5 (B+) \\
P09 & Male & 77.5 (B+) \\
P10 & Male & 87.5 (A+) \\
P11 & Male & 100 (A+) \\
P12 & Male & 70 (C) \\
P13 & Male & 77.5 (B+) \\ \hline
\end{tabular}
\label{tab:sus}
\end{center}
\end{table}

\subsection{Usability Rest}
In addition to the main result, we asked the participants to answer the System usability scale (SUS) test~\cite{brooke1996sus}.
Table~\ref{tab:sus} summarizes the SUS scores of each participant.
The average score and its 95\% confidence interval among all the participants were 75.4 $\pm$ 8.6. 
Based on acceptable ranges~\cite{bangor2009determining}, 8 out of 12 participants evaluated as acceptable while one participant (P04) evaluated as unacceptable.
Low-scored participants mainly pointed out difficulty in browsing the object timeline (discussed in 6.4).

\subsection{Comfort on Neck-mounted Camera}
We asked the question, ``How comfortable was the neck-mounted camera?'' on a seven-point scale (unpleasant\,=\,1, comfortable\,=\,7).
The participants reported slightly positive feedback on average regarding comfort of camera (mean and 95\% CI: 4.6 $\pm$1.2).
Some preferred attaching cameras on their glasses instead of their necks.

\subsection{Object Discovery Examples}
Figure~\ref{fig:timeline_01b} and \ref{fig:timeline_09b} show additional examples of the obtained hand-held object timeline.

\bfigure
\centerline{\includegraphics[width=1.0\linewidth]{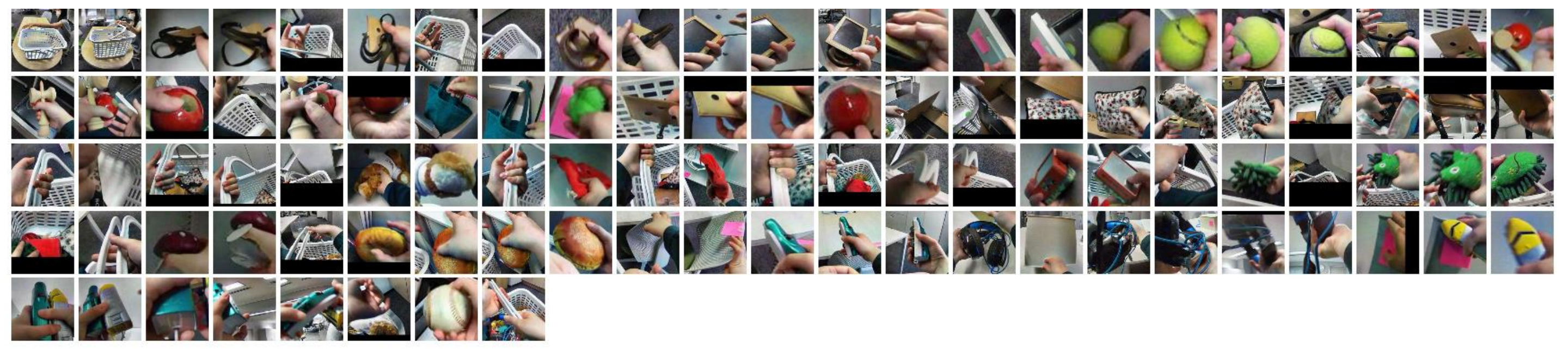}}
\caption{More results of hand-held object timeline (localization rate=0.9375, \#clusters=96).}
\label{fig:timeline_01b}
\efigure

\bfigure
\centerline{\includegraphics[width=1.0\linewidth]{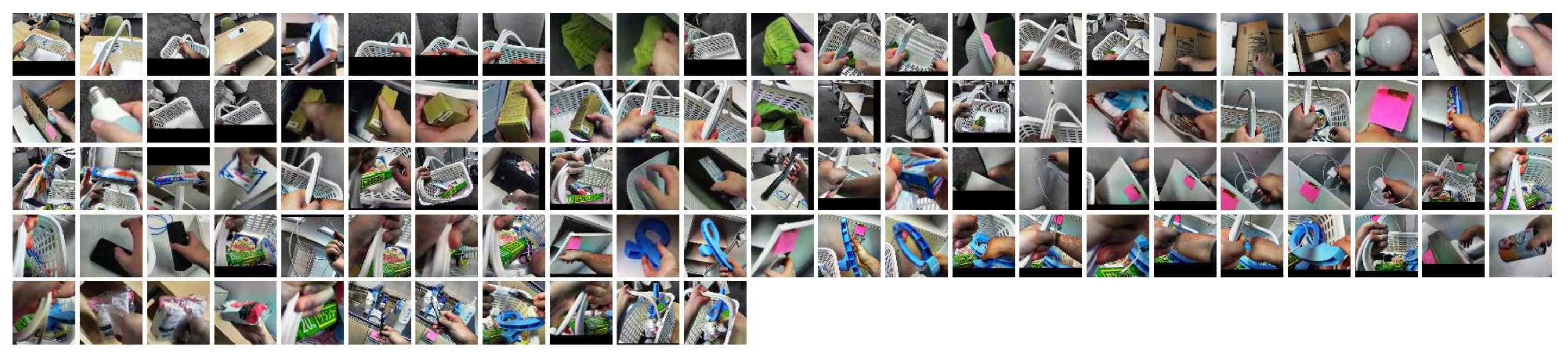}}
\caption{More results of hand-held object timeline (localization rate=0.75, \#clusters=99).}
\label{fig:timeline_09b}
\efigure

\end{document}